\documentclass[aps, prb, twocolumn]{revtex4}
\usepackage{amssymb}
\usepackage{graphicx}
\usepackage{hyperref}

\begin{document}

\title{Phase Relations in the Li$_2$O-V$_2$O$_3$-V$_2$O$_5$ System at 700 $^\circ$C: Correlations with Magnetic Defect Concentration in Heavy Fermion LiV$_2$O$_4$}
\author{S. Das}
\author{X. Ma}
\altaffiliation[Present address: ]{Chemical Physics Program, Institute for Physical Science and Technology, University of Maryland, College Park, MD 20742.}
\author{X. Zong}
\author{A. Niazi}
\author{D. C. Johnston}
\affiliation{Ames Laboratory and Department of Physics and Astronomy, Iowa State University, Ames, Iowa 50011}

\date{\today}

\begin{abstract}

The phase relations in the ${\rm Li_2O}$-${\rm V_2O_3}$-${\rm V_2O_5}$ ternary system at 700 $^\circ$C for compositions in equilibrium with ${\rm LiV_2O_4}$ are reported. This study clarified the synthesis conditions under which low and high magnetic defect concentrations can be obtained within the spinel structure of ${\rm LiV_2O_4}$. We confirmed that the ${\rm LiV_2O_4}$ phase can be obtained containing low (0.006 mol\%) to high ( 0.83 mol\%) magnetic defect concentrations $n{\rm _{defect}}$ and with consistently high magnetic defect spin $S$ values between 3 and 6.5.  The high $n{\rm _{defect}}$ values were obtained in the ${\rm LiV_2O_4}$ phase in equilibrium with ${\rm V_2O_3}$, ${\rm Li_3VO_4}$, or ${\rm LiVO_2}$ and the low values in the ${\rm LiV_2O_4}$ phase in equilibrium with ${\rm V_3O_5}$. A model is suggested to explain this correlation. 

\end{abstract}

\maketitle

\section{\label{intro}Introduction}

Heavy fermion (HF) behavior has mostly been seen in \textit{f}-electron metals. Such compounds are called heavy fermions because in these materials the current carriers behave as if they have a large mass ($\sim$ 10$^{2}$--$10^{3}$ times the free electron mass). ${\rm LiV_2O_4}$, first synthesized by Reuter and Jaskowsky,\cite{reuter} is one of the few $d$-electron compounds to show HF behaviour at low temperatures.\cite{Kondo1997, Johnston2000} ${\rm LiV_2O_4}$ has the face-centered-cubic spinel structure with the space group \textit{Fd$\overline{3}$m}. The V atoms are coordinated by six O atoms in slightly distorted octahedron. The Li atoms are coordinated with four O atoms in a tetrahedron. The Li atoms are located in the gaps between chains of the ${\rm VO_6}$ edge-sharing octahedra. From NMR measurements done on ${\rm LiV_2O_4}$ samples it has been found that for magnetically pure samples the $^{7}$Li nuclear spin-lattice relaxation rate ${1/T_1}$ is proportional to temperature \textit{T} at low temperatures (the Korringa law) which is typical for Fermi liquids.\cite{Kondo1997, Mahajan1998, Fujiwara2004} However for samples which contain magnetic defects within the spinel structure, the relaxation rate has a peak at $\sim1$ K and also shows other signatures which do not agree with the behavior of Fermi liquids.\cite{Johnston2005} The occurrence of magnetic defects is easily seen as a low-$T$ Curie-like upturn in the magnetic susceptibility rather than becoming nearly independant of $T$ below $\sim$ 10 K as observed for the intrinsic behavior.\cite{Kondo1999} The mechanism for the formation of the magnetic defects is not known yet.

Previously, polycrystalline samples of ${\rm LiV_2O_4}$ had been prepared from the starting materials ${\rm Li_2CO_3}$, ${\rm V_2O_3}$ and ${\rm V_2O_5}$ at 700 $^\circ$C.  Typically, the samples contain a concentration of magnetic defects $n$${\rm _{defect}}$ within the structure of the spinel phase, ranging from $\lesssim$ 0.01  to 0.7 mol\%.\cite{Kondo1999} Magnetization $M$ versus applied magnetic field $H$ measurements at low $T$ were carried out to estimate $n{\rm _{defect}}$ and the defect spin $S$${\rm _{defect}}$. Low concentrations of defects were found in samples of ${\rm LiV_2O_4}$ containing small amounts of ${\rm V_3O_5}$ impurity phase  while high defect concentrations were found in samples containing ${\rm V_2O_3}$ impurity phase.\cite{Kondo1999} Though the reason behind this correlation is not known yet, these results pointed towards a controllable way to vary the magnetic defect concentration within the spinel structure. However, it was not clear that the above-noted ${\rm V_2O_3}$ and ${\rm V_3O_5}$ impurity phases were in equilibrium with the ${\rm LiV_2O_4}$ spinel phase at 700 $^\circ$C. In addition, it was unknown (in Ref. [7]) how the magnetic defect concentration in the spinel phase ${\rm LiV_2O_4}$ varied if other impurity phases were present. To help resolve these questions, we report here the phase relations in the ${\rm Li_2O}$--${\rm V_2O_3}$--${\rm V_2O_5}$ system at 700  $^\circ$C, in the vicinity of the composition ${\rm LiV_2O_4}$, and report the magnetic properties of a selection of such compositions. 

There have been some studies of the ${\rm Li_2O}$--${\rm V_2O_5}$ system which revealed the existence of three phases in the system, namely ${\rm LiVO_3}$, ${\rm Li_3VO_4}$ and ${\rm LiV_3O_8}$. Reisman et al.\cite{Reisman} reported the existence of the congruently melting phases ${\rm LiVO_3}$ (reported as ${\rm Li_2O}$$\cdot$${\rm V_2O_5}$) and ${\rm Li_3VO_4}$ (reported as ${\rm 3Li_2O}$$\cdot$${\rm V_2O_5}$) with melting points 616 $^\circ$C and 1152 $^\circ$C, respectively. ${\rm LiV_3O_8}$ has been reported to be both congruently melting and incongruently melting.\cite{Kohmuller,Reisman,Wickham} Manthiram et al.\cite{manthiram} reported that Li$_{1-x}$VO$_2$ is single phase in the compositional range $0 \le x \le 0.3$ at 650~$^\circ$C\@. ${\rm LiV_2O_4}$ was reported to exist in equilibrium with the compounds ${\rm VO_2}$ and Li$_{1-x}$VO$_2$ from room temperature to 1000 $^\circ$C by Goodenough et al.\cite{goodenough} The lithium vanadium oxide system Li$_x$V$_2$O$_5$, also known as the lithium vanadium bronze phase, was reported to exist in a number of single-phase regions for $0<x<1$ and temperature $T$ $<$ 500 $^\circ$C\@.\cite{Murphy}

The ${\rm V_2O_3}$--${\rm V_2O_5}$ binary system has been extensively studied and a large number of phases have been reported. Hoschek and Klemm\cite{Hoschek1939} first studied the system and suggested the presence of the phase ${\rm V_2O_3}$, the $\beta$-phase (${\rm VO_{1.65}}$--${\rm VO_{1.80}}$), the $\alpha$-phase (${\rm VO_{1.80}}$--${\rm VO_2}$), and the $\alpha^\prime$-phase (${\rm VO_{2.09}}$--${\rm VO_{2.23}}$). Andersson\cite{AnderssonG} reported phases with general formula V$_n$O$_{2n-1}$ with 3 $\leq n<9$. Additional phases reported in this system are ${\rm V_9O_{17}}$ and ${\rm V_{10}O_{19}}$\cite{Kuwamoto}. The phases with general formula V$_n$O$_{2n-1}$ with 3 $\leq n\leq9$ are called the Magn\'eli phases.\cite{Magneli} The triclinic structure of the Magn\'eli phases have been reported.\cite{AnderssonG,AnderssonS,Horiuchi,Kuwamoto} The other V-O phases existing between ${\rm VO_2}$ and ${\rm V_2O_5}$ are ${\rm V_6O_{13}},$\cite{Aebi,AnderssonG} ${\rm V_4O_9}$\cite{Wilhelmi} and ${\rm V_3O_7}$.\cite{Tudo,Kosuge} Combined with the work by Kachi and Roy\cite{kachi}, Kosuge\cite{Kosuge} proposed a phase diagram of the ${\rm V_2O_3}$-${\rm V_2O_5}$ system in the temperature-composition plane extending from room temperature to 1200~$^\circ$C showing high melting points ($>1200$ $^\circ$C) for V-O phases existing between ${\rm V_2O_3}$ and ${\rm VO_2}$, low melting points ($\lesssim700$ $^\circ$C) for V-O phases existing between ${\rm VO_2}$ and ${\rm V_2O_5}$ and also the homogeneity ranges of all the phases existing between ${\rm V_2O_3}$ and ${\rm V_2O_5}$.

\section{\label{exp}experimental details}

Our samples were prepared by conventional solid state reaction as described by Kondo et al.\cite{Kondo1999} The starting materials were ${\rm Li_2CO_3}$ (99.995\%, Alfa Aesar), ${\rm V_2O_5}$ (99.995\%, M V Laboratories Inc.) and ${\rm V_2O_3}$ (99.999\%, M V Laboratories Inc.). The samples were made in two stages. First a (Li$_2$O)$_x$(V$_2$O$_5$)$_y$ precursor was made by thoroughly mixing appropriate amounts of ${\rm Li_2CO_3}$ and ${\rm V_2O_5}$, pressing into a pellet and then heating in a tube furnace under oxygen flow at 525 $^\circ$C until the expected weight loss occured due to the loss of ${\rm CO_2}$ from ${\rm Li_2CO_3}$. The precursor pellet was then crushed and the appropriate amount of ${\rm V_2O_3}$ was added and mixed thoroughly inside a helium-filled glove box. The precursor-${\rm V_2O_3}$ mixture was then again pressed into a pellet, wrapped in a platinum foil, sealed in a quartz tube under vacuum and then heated at 700~$^\circ$C for about ten days. The samples were taken out of the furnace and air-cooled to room temperature. The different phases present in the samples were identified from X-ray diffraction patterns at room temperature obtained using a Rikagu Geigerflex diffractometer with a curved graphite crystal monochromator. The diffraction patterns were matched with known phases from the JCPDS\cite{jcpds} database using the JADE 7 program.\cite{jade} The samples were repeatedly ground and heated until the X-ray patterns did not show any change to ensure that the samples were in thermal equilibrium at 700 $^\circ$C. The magnetization $M$${\rm _{obs}}$ measurements were done on the samples using a Quantum Design Superconducting Quantum Interference Device (SQUID) magnetometer over the temperature $T$ range 1.8~K -- 350~K and applied magnetic field $H$ range 0.001~T -- 5.5~T.

\section{\label{results}results and analysis}

\subsection{\label{Phase relations}Phase Relations at 700 $^\circ$C}

The phase relations for phases in equilibrium with ${\rm LiV_2O_4}$ at 700~$^\circ$C are shown in  Fig.~1. The black triangles represent the crystalline phases which exist singly in equilibrium at 700~$^\circ$C. The solid dots represent the compositions of our samples from which the phase relations were determined. The solid straight lines connecting the phases are the tie lines. From a large number of samples synthesized at the nominal stoichiometric composition ${\rm LiV_2O_4}$, it has been found that ${\rm LiV_2O_4}$ is a ``line compound'', i.e, this compound has an extremely small ($\lesssim$ 1 at.\%) homogeneity range.  This situation is very different from the large homogeneity range $0 \le x \le 1/3$ in the similiar spinel phase Li[Li$_x$Ti$_{2-x}$]O$_4$.\cite{Johnston} According to the study of Li$_{1-x}$VO$_2$ by Goodenough et al.\cite{goodenough} mentioned above, there is a tie line between ${\rm LiV_2O_4}$ and ${\rm LiVO_2}$ at 700 $^\circ$C, consistent with our results. However, our results conflict with their finding of a tie line between ${\rm LiV_2O_4}$ and ${\rm VO_2}$. In particular, the observed tie line in Fig. 1 between ${\rm V_4O_7}$ and ${\rm Li_3VO_4}$ precludes a tie line between ${\rm LiV_2O_4}$ and ${\rm VO_2}$ because the latter would have to cross the former which is not allowed.

\begin{figure}[t]
\includegraphics[width=3.5in]{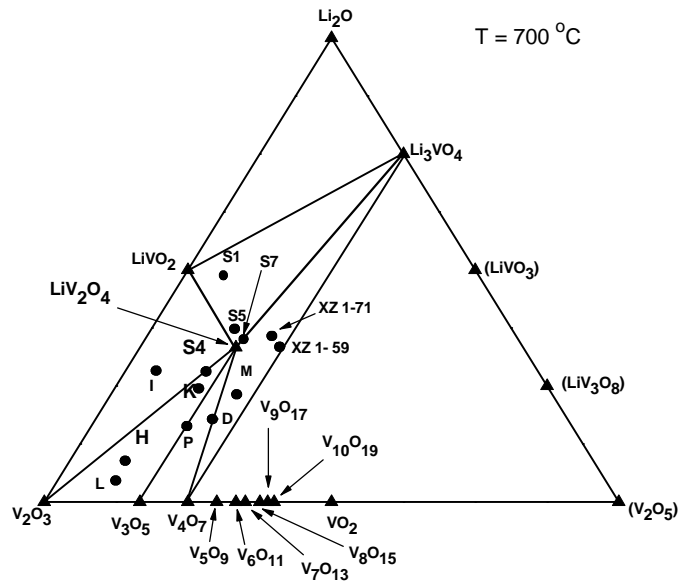}
\label{ternary diagram}
\caption{Phase relations in the ${\rm Li_2O}$-${\rm V_2O_3}$-${\rm V_2O_5}$ system at 700 $^\circ$C for phases in equilibrium with the ${\rm LiV_2O_4}$ spinel phase. The triangles represent the crystalline phases which exist singly in equilibrium at 700 $^\circ$C. The dots represent the compositions of our samples from which the phase relations were determined. The solid straight lines connecting the phases are the tie lines. The compounds in parentheses melt below 700 $^\circ$C.}
\end{figure}  

\subsection{\label{magnetic}Magnetic measurements}

\subsubsection{\label{magnetic}Magnetic susceptibility measurements}

Here we present the magnetic susceptibility $\chi$ versus temperature $T$ for some of our samples of ${\rm LiV_2O_4}$ containing small amounts ($\lesssim 2$ wt\%) of impurity phases. Based on the X-ray diffraction patterns, the impurity phases present in the samples are ${\rm V_2O_3}$ in sample 5A, ${\rm V_3O_5}$ in sample 8 , ${\rm LiVO_2}$ in sample 5B, and ${\rm Li_3VO_4}$ in sample S7 as shown in Table I. Sample 6B was the crystallographically purest sample synthesized and the X-ray diffraction pattern did not reveal any impurity phases. Figures 2 and 3 show expanded X-ray diffraction patterns of these samples. 

\begin{figure}[t]
\includegraphics[width=3in]{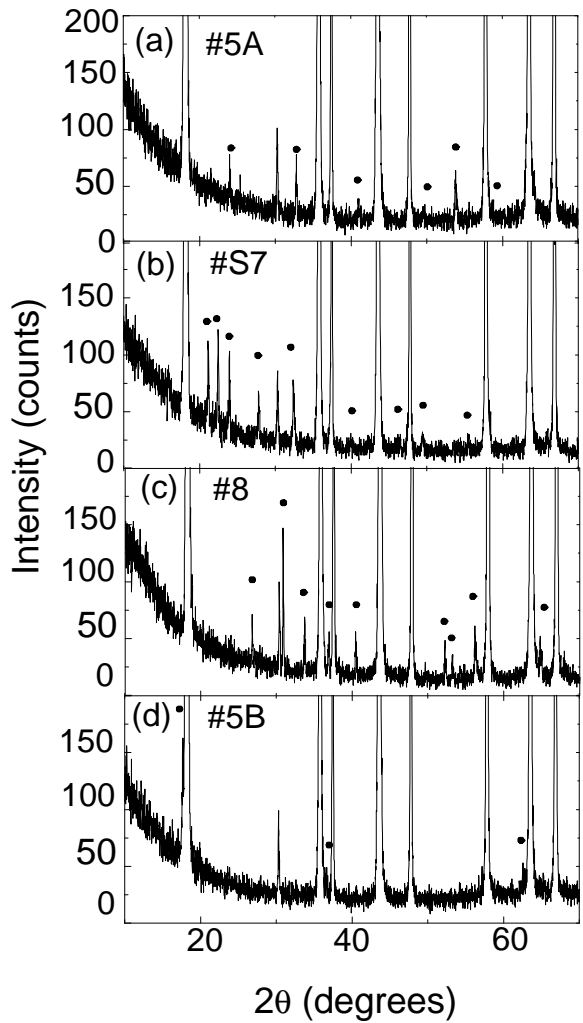}
\label{all xrays}
\caption{Expanded X-ray diffraction patterns of samples with compositions near ${\rm LiV_2O_4}$. The impurity phase peaks are marked by solid circles. (a) Sample 5A has ${\rm V_2O_3}$ impurity phase. (b) Sample S7 has ${\rm Li_3VO_4}$ impurity phase. (c) Sample 8 has ${\rm V_3O_5}$ impurity phase. (d) Sample 5B has ${\rm LiVO_2}$ impurity phase.}
\end{figure}  

\begin{figure}[t]
\includegraphics[width=3in]{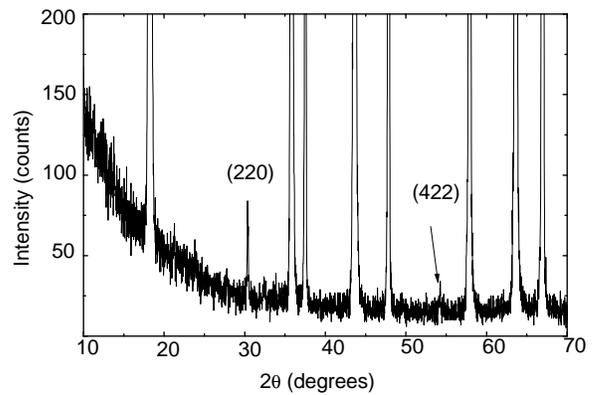}
\label{xray 6B}
\caption{Expanded X-ray diffraction pattern of the ${\rm LiV_2O_4}$ sample 6B. The two indexed peaks are of the ${\rm LiV_2O_4}$ spinel phase. There are no observable impurity phases present. }
\end{figure}

The observed magnetic susceptibility $\chi$${\rm _{obs}}$ versus $T$ plots from $T$ = 1.8 K to 350 K at magnetic field $H$ = 1 T for the five samples are shown in Fig. 4 where $\chi$${\rm _{obs}}$ $\equiv$ $M$${\rm _{obs}}$/$H$. It can be clearly seen that the dependence of $\chi_{\rm {obs}}$ on $T$ for the five samples is similar Curie-Weiss like for $T$ $>$ 50 K\@. However for $T$ $<$ 50 K the dependence is strikingly different. Sample 8 containing ${\rm V_3O_5}$ impurity phase shows a broad peak at $T$ $\approx$ 20 K, which is characteristic of the intrinsic behavior of magnetically pure ${\rm LiV_2O_4}$.\cite{Kondo1999} Sample 6B which is crystallographically pure also shows a broad peak but it is masked by a Curie-like upturn at $T$ $<$ 10 K. For sample 5A containing ${\rm V_2O_3}$, S7 containing ${\rm Li_3VO_4}$, and 5B containing ${\rm LiVO_2}$ as impurity phases, the broad peak is totally masked by Curie contributions. 

To interpret the origin of the Curie-like low-$T$ contributions to $\chi(T)$ of these samples, it is important to consider the potential contributions of the impurity phases to this term.  ${\rm V_3O_5}$ orders antiferromagnetically with its susceptibility showing a very broad maximum between $T$ = 120 K and 130 K\cite{nagata,ueda} which is much higher than its N\'{e}el temperature $T_{\rm N} = 75.5$~K measured by Griffing.\cite{griffing} The susceptibility for $T < T_{\rm N}$ decreases with decreasing $T$,  has a value $< 2\times 10^{-5}$ cm$^3$/mol at the lowest temperatures, and shows no evidence for a Curie-like term. ${\rm V_2O_3}$ has a Curie-Weiss-like behaviour for $T > 170$ K where it is also metallic. Below 170 K it orders antiferromagnetically at a metal to insulator transition and the susceptibility remains constant at about $5\times 10^{-4}$ cm$^{3}$/mol down to $T$ $\sim$ 80 K. For $T$ $<$ 80 K, the susceptibility decreases with decreasing $T$ with no sign of a Curie-like upturn.\cite{ueda, Kikuchi} The susceptibility of V$_{2-y}$O$_3$ shows a peak at low $T$ ($\sim 10$ K) as it undergoes antiferromagnetic ordering at around 10 K with no evidence for a Curie-like term at lower $T$.\cite{ueda} ${\rm Li_3VO_4}$ is nonmagnetic since the vanadium atom is in the +5 oxidation state. The only impurity phase exhibiting a low-temperature Curie-like contribution to its susceptibility is Li$_{1-x}$VO$_2$, which shows a Curie-like upturn at $T < 50$ K due to Li deficiency of about 5\%.\cite{Tian,Onoda} However, the amounts of impurity phases in our ${\rm LiV_2O_4}$ samples are small ($<$ 2 wt\%). Assuming that $x$ = 0.05 in Li$_{1-x}$VO$_2$ impurity phase,\cite{Tian} where each Li vacancy induces a V$^{+4}$ ($S$ = 1/2) defect in that phase, one obtains a Curie constant of $\simeq4\times10^{-4}$ cm$^3$ K/mol, which is far smaller than observed ($\sim0.1$ cm$^3$ K/mol) in our sample 5B having Li$_{1-x}$VO$_2$ impurity phase. Thus we can conclude that the Curie-like upturn in the susceptibility of nearly single-phase LiV$_2$O$_4$ arises from magnetic defects within the spinel structure of this compound and not from impurity phases, which confirms the previous  conclusion of Ref. [7].

\begin{figure}[t]
\includegraphics[width=3.3in]{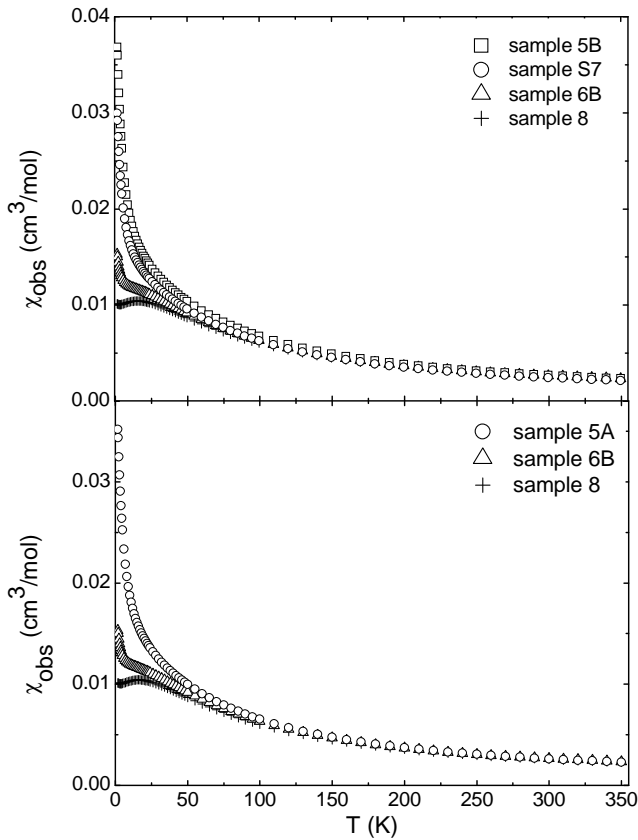}
\label{magnetic susceptibility curves}
\caption{Observed magnetic susceptibility $\chi$${\rm _{obs}}$ versus temperature $T$ at $H=1$ T for several samples in Table I that are nearly single-phase ${\rm LiV_2O_4}$.}
\end{figure}

\subsubsection{\label{magnetic}Isothermal magnetization measurements}

The observed magnetization $M{\rm _{obs}}$ versus applied magnetic field $H$ isothermal measurements were done at different temperatures between 1.8 K and 350 K with $H$ varying from 0.001 T to 5.5 T. However, to find $n{\rm _{defect}}$ only the low $T$ (1.8 K, 2.5 K, 3 K and 5 K) isotherms were used. The $M{\rm _{obs}}$ versus $H$ curves for different samples at 1.8 K are shown in Fig. 5. The samples showing a Curie-like upturn in the susceptibility show a negative curvature in their $M{\rm _{obs}}$ versus $H$ curves, whereas the samples having a very small Curie-like upturn in the susceptibility show a hardly observable curvature. This correlation shows that the Curie contribution to the susceptibility is due to field saturable (paramagnetic) defects. The values of the defect concentrations and the values of the defect spins for different samples were determined according to the analysis done by Kondo et al.\cite{Kondo1999} The observed molar magnetization $M$${\rm _{obs}}$ isotherms at low temperatures ($T$ $\le$ 5 K) for each sample were simultaneously fitted by 
 
\begin{equation}
M{\rm _{obs}} = \chi H + n_{\rm defect} N{\rm _A}g{\rm _{defect}}\mu{\rm _B} S{\rm _{defect}} B_{S}(x)\, , 
\label{fiteq}
\end{equation}
where ${n_{\rm defect}}$ is the concentration of the magnetic defects, $N{\rm _A}$ Avogadro's number, $g{\rm _{defect}}$ the $g$-factor of the defect spins which was fixed to 2 (the detailed reasoning behind this is given in Ref. [7]), $S{\rm _{defect}}$ the spin of the defects, $B _S(x)$ the Brillouin function, and $\chi$ the intrinsic susceptibility of ${\rm LiV_2O_4}$ spinel phase. The argument of the Brillouin function $B _S(x)$  is $x$~=~$g{\rm _{defect}}$$\mu{\rm _B}$$S{\rm _{defect}}$$H$/[k${\rm _B}$($T$$-$$\theta{\rm _{defect}}$)] where $\theta{\rm _{defect}}$ is the Weiss temperature. 
The four fitting parameters $\chi$, $n_{\rm defect}$, $S{\rm _{defect}}$ and $\theta{\rm _{defect}}$ for each sample are listed in Table I. Since the parameters $n{\rm _{defect}}$ and $S{\rm _{defect}}$ are strongly correlated in the fits, the products of these are also listed in Table I. 

The grain sizes of our samples were studied using a scanning electron microscope (SEM). The SEM pictures of some of our samples are shown in Fig.\ 6. As seen from the figure,  the grain sizes are 1 -- 10 $\mu$m, and from Table I there is no evident correlation between the sample grain sizes and the magnetic defect concentrations.

\begin{figure}[t]
\includegraphics[width=3in]{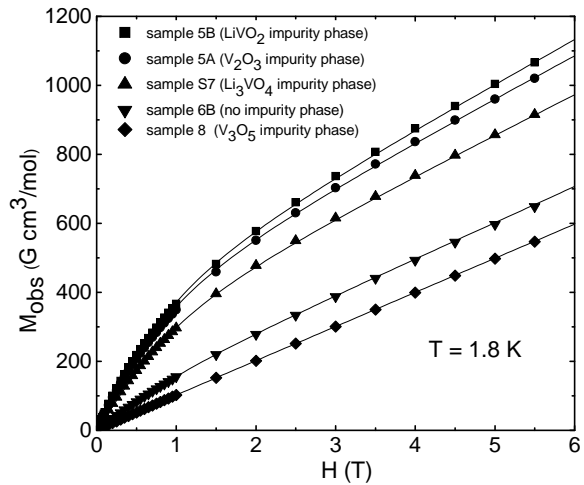}
\label{mh all}
\caption{$M$${\rm _{obs}}$ versus $H$ isotherms of four samples at 1.8 K. The curves passing through the data points are fits by Eq. (1) with the values of the parameters given in Table I.}
\end{figure}

\begin{figure}[b]
\includegraphics[width=3.5in]{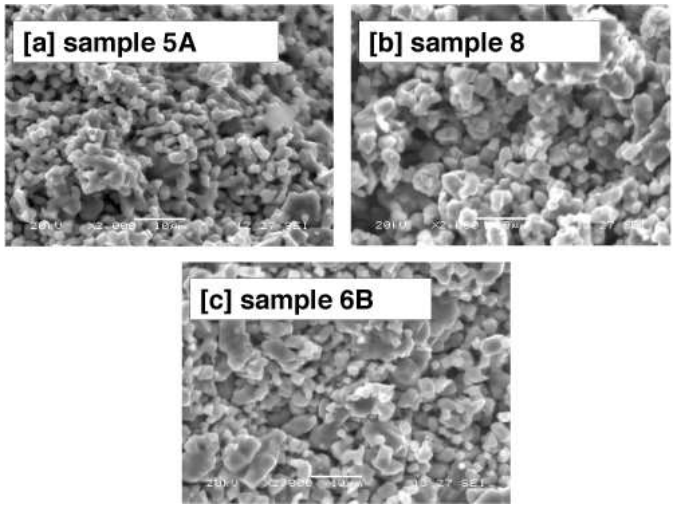}
\label{sem all}
\caption{SEM pictures of our ${\rm LiV_2O_4}$ powder samples. No evident correlation between the grain sizes and the defect concentrations was found. The bars at the bottom of each picture are 10 $\mu$m long. The grain sizes are in the range 1 to 10 $\mu$m.}
\end{figure}

\begin{figure}[t]
\includegraphics[width=2.75in]{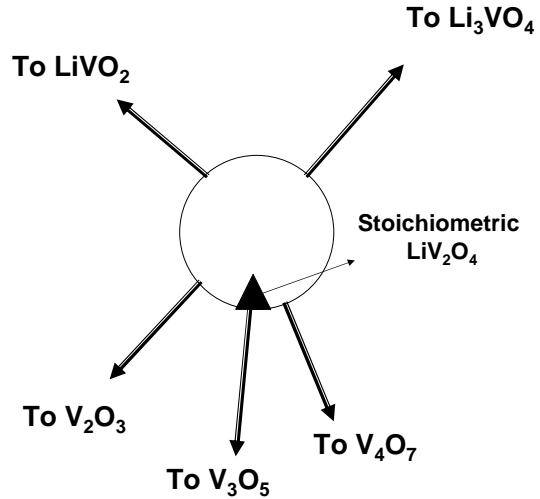}
\label{probable model}
\caption{Suggested model for the mechanism of the crystal and magnetic defect formation in ${\rm LiV_2O_4}$. The figure shows an enlarged region around ${\rm LiV_2O_4}$ in the phase relation picture (Fig.\ 1) where the circle represents a possible small homogeneity range of the spinel phase and the filled triangle is stoichiometric ${\rm LiV_2O_4}$.}\
\end{figure}

\begin{table*}
\caption{Results of the analyses of the $M_{\rm _{obs}}(H, T)$. The error in the last digit of a parameter is given in parentheses.}
\begin{ruledtabular}
\begin{tabular}{lllllll}
Sample no & Impurity & $\chi$ (cm$^3$/mol) & $n{\rm _{defect}}$ (mol\%) & $S$${\rm _{defect}}$ & $\theta{\rm _{defect}}$ (K) & $n{\rm _{defect}}$$S$${\rm _{defect}}$ (mol\%)\\
\hline
5A & ${\rm V_2O_3}$ & 0.0123(1) & 0.77(3) & 4.0(1) & $-$0.70(13) & 3.08(13)\\
S7 & ${\rm Li_3VO_4}$ & 0.0115(1) & 0.67(2) & 3.7(1) & $-$0.59(9) & 2.52(8)\\
8  & ${\rm V_3O_5}$ & 0.0098(1) & 0.0067(28) & 6.3(27) & $-$1.0(10) & 0.04(18)\\
5B & ${\rm LiVO_2}$ & 0.0127(2) & 0.83(3) & 3.9(1) & $-$0.65(12) & 3.29(13)\\
6B & no impurity & 0.0104(1) & 0.21(1) & 3.5(2) & $-$0.75(13) & 0.73(4)\\
\\
\\
\end{tabular}
\end{ruledtabular}
\label{table1}
\end{table*}

\section{\label{model}Suggested model}

The reason behind the correlation between the presence of the Li-V-O  and V-O phases and the variation of the magnetic defect concentration in ${\rm LiV_2O_4}$ is not known yet. We speculate that this is due to the formation of vacancies and/or interstitials in the spinel structure due to the variation of the sample composition from the ideal stoichiometry. A possible model is shown in Fig.\ 7. The black triangle is stoichiometric ${\rm LiV_2O_4}$ while the circular region is a small ($\lesssim1$ at.\%) homogeneity range of ${\rm LiV_2O_4}$. Based on this model, the ${\rm LiV_2O_4}$ phase in the samples having ${\rm V_3O_5}$ impurity phase are very close to the ideal stoichiometric ${\rm LiV_2O_4}$, the magnetic susceptibility is the intrinsic susceptibility for the ideal stoichiometric spinel phase and the magnetic defect concentration is very small. The composition of the spinel phase in samples having ${\rm V_2O_3}$, ${\rm Li_3VO_4}$ or ${\rm LiVO_2}$ as impurity phases deviates from the ideal stoichiometry as can be seen in the figure. This variation from the ideal stoichiometry would cause the above vacancies and/or interstitial defects to form which in turn cause the formation of paramagnetic defects. The samples having chemical composition different from the black solid triangle (i.e.\ the ideal stoichiometric composition) but within the circular region will be by definition single phase ${\rm LiV_2O_4}$ but not having the ideal stoichiometry. Thus some samples of ${\rm LiV_2O_4}$ will have magnetic defects even if there are no impurity phases in them which might be the case for our sample 6B and also samples 3 and 7 studied by Kondo et al.,\cite{Kondo1999} where some samples were essentially impurity free but still had a strong Curie contribution in their susceptibility.

\section{\label{conclusion}conclusion}

In this paper we have reported the phase relations in the ${\rm Li_2O}$-${\rm V_2O_3}$-${\rm V_2O_5}$ system at 700 $^\circ$C for compositions in equilibrium with ${\rm LiV_2O_4}$. This study helped us to determine the synthesis conditions under which polycrystalline samples of ${\rm LiV_2O_4}$ could be prepared with variable magnetic defect concentrations ranging from $n$${\rm _{defect}}$ = 0.006 to 0.83 mol\%. High magnetic defect concentrations were found in samples containing ${\rm V_2O_3}$, ${\rm Li_3VO_4}$, or ${\rm LiVO_2}$ impurity phases while the samples containing ${\rm V_3O_5}$ impurity phase had low defect concentration. We suggested a possible model which might explain this correlation. Our work shows how to systematically and controllably synthesize ${\rm LiV_2O_4}$ samples with variable magnetic defect concentrations within the spinel structure. The results should be helpful to other researchers synthesizing samples for study of the physical properties of this system.

\begin{acknowledgments}
Ames Laboratory is operated for the U. S.\ Department of Energy by Iowa State University under Contract No.\ W-7405-Eng-82. This work was supported by the Director for Energy Research, Office of Basic Energy Sciences.
\end{acknowledgments}

\end{document}